\documentclass[twocolumn,amsmath,amssymb,pra]{revtex4}
\usepackage{graphicx}
\usepackage{bm}
\usepackage[tight]{subfigure}
\newcounter{one}
\begin{document}

\title{Efficient generation of highly squeezed light and second harmonic wave with periodically poled MgO:LiNbO$_{\mathrm{3}}$}

\author
{Genta Masada$^{1,2}$, Tsuyoshi Suzudo$^{3}$, Yasuhiro Satoh$^{3}$, Hideki Ishizuki$^{4}$, Takunori Taira$^{4}$, and Akira Furusawa$^{1}$}

\affiliation{$^{1}$Department of Applied Physics and Quantum-Phase Electronics Center, School of Engineering, The University of Tokyo, 7-3-1 Hongo, Bunkyo-ku, Tokyo 113-8656, Japan \\
$^{2}$Tamagawa University Research Institute, 6-1-1 Tamagawagakuen, Machida-city, Tokyo 194-8610, Japan \\
$^{3}$Ricoh Company, Ltd., 5-10 Yokarakami, Kumanodo, Takadate, Natori-city, Miyagi 981-1241, Japan \\
$^{4}$Institute for Molecular Science, 38 Nishigonaka, Myodaiji, Okazaki-city, Aichi 444-8585, Japan}

\begin{abstract}
We report on effective generation of continuous-wave squeezed light and second harmonics with a periodically poled MgO:LiNbO$_{\mathrm{3}}$ (PPMgLN) crystal which enables us to utilize the large nonlinear optical coefficient $d_{\mathrm{33}}$.
We achieved the squeezing level of $-7.60 \pm 0.15$dB at 860 nm by utilizing a subthreshol optical parametric oscillator with a PPMgLN crystal.
We also generated 400 mW of second harmonics at 430 nm from 570 mW of fundamental waves with 70\% of conversion efficiency by using a PPMgLN crystal inside an external cavity.
 
\end{abstract}


\maketitle

One of the attractive application of squeezed light is quantum information processing with continuous variables~\cite{Braunstein05}. 
Quadrature squeezed vacuum states are applied to realize quantum teleportation which is a fundamental protocol in quantum information processing~\cite{Furusawa98}. 
The fidelity of such protocols is limited directly by the squeezing level~\cite{Suzuki06}. 
So it is important to generate highly squeezed light to achieve a better performance.
A typical method to generate highly squeezed light is utilization of a subthreshold optical parametric oscillator (OPO) which includes a nonlinear optical medium. 
Over the past few decades a considerable number of the experiments have been performed to generate highly squeezed light as shown in Table.~\ref{table}.
The oscillation threshold of pump power $P_{th}$ and the escape efficiency $\rho$ for OPOs are important factors to generate highly squeezed light, $P_{th}=(T+L)^2/4E_{NL}$ and $\rho=T/(T+L)$ respectively, where ${E}_{NL}$ is the effective nonlinearity of optical mediums, $T$ is the transmittance of output coupler, and $L$ is the intracavity loss. Polzik,\textit{et al}. achieved $-6.0 \pm0.3$dB squeezing level at 852 nm with a bow tie configuration of the OPO including a KNbO$_{\mathrm{3}}$ (KN) crystal as a nonlinear medium~\cite{Polzik92}. 
A major factor of degradation of the observed squeezing level was intracavity losses caused by blue light induced infrared absorption (BLIIRA) in a KN crystal. 
Suzuki,\textit{et al.} and Takeno,\textit{et al.} achieved $-7.2 \pm0.14$dB~\cite{Suzuki06} and $-9.01 \pm0.14$dB~\cite{Takeno07} at 860 nm respectively with a periodically poled KTiOPO$_{\mathrm{4}}$ (PPKTP) crystal which has rather low losses without BLIIRA. 
Recently Vahlbruch,\textit{et al.} utilized a monolithic OPO cavity by a MgO 7mol\% doped LiNbO$_{\mathrm{3}}$ (MgLN) single crystal in order to reduce losses caused by extra optical elements. 
They succeeded in measuring $-10.12 \pm0.15$dB of squeezing at 1064 nm ~\cite{Vahlbruch08}.
However relatively low nonlinear optical coefficient $d_{\mathrm{31}}$ of MgLN crystal causes an increase of the oscillation threshold.
From this passage, they placed more considerable emphasis on the reduction of the intracavity losses than nonlinearity of the medium to improve the escape efficiency.  

\begin{table*}[!htb]
\centering
\begin{tabular}{|l|c|c|c|c|}
 \hline
 Nonlinear optical medium                            & KN                   & PPKTP           & MgLN                     & PPMgLN  \\
 Effective nonlinear optical coefficient $d_{\textit{eff}}$ (pm/V)~\cite{Shoji97}& $d_{\mathrm{31}}$=11.0    & $\frac{2}{\pi}d_{\mathrm{33}}$=10.6   & $d_{\mathrm{31}}\approx 4.4$  & $\frac{2}{\pi}d_{\mathrm{33}}$=18.1  \\ \hline
 Squeezing level (dB)       & $-6.0\pm0.3$~\cite{Polzik92}         & $-9.01\pm0.14$~\cite{Suzuki06,Takeno07}    & $-10.12\pm0.15$~\cite{Vahlbruch08}   & $-7.60\pm0.15$ \\
 Oscillation threshold $P_{th}$ (mW)                 & 250                  & 180             & 744$^{**}$               & 377 (110) \\
 Escape efficiency $\rho$                            & 0.875                & 0.97            & 0.994                    & 0.925 \\
 Effective nonlinearity ${E}_{NL}$ ($\mathrm{W}^{-1}$)  & 0.016             & 0.023           & 0.0049$^{**}$            &0.043 \\
 Transmittance of output coupler $T$                 & 0.105                & 0.123           & 0.12                     & 0.21 (0.113) \\
 Intracavity loss $L$                                & 0.015$^*$            & 0.0038          & 0.0007                   & 0.017 \\ \hline

\end{tabular}
\caption{
Summary of previous squeezing experiments. Nonlinear optical coefficient of MgLN is the value of MgO 5\% doped LiNbO$_{\mathrm{3}}$ at 1064 nm and others are values at 852 nm. 
The values indicated with $^*$ and $^{**}$ are estimated with experimental parameter described in references~\cite{Polzik92,Vahlbruch08} respectively. 
}
\label{table}
\end{table*}

\begin{figure}
\centering
\subfigure[]{
\includegraphics[width=3.5cm,clip]{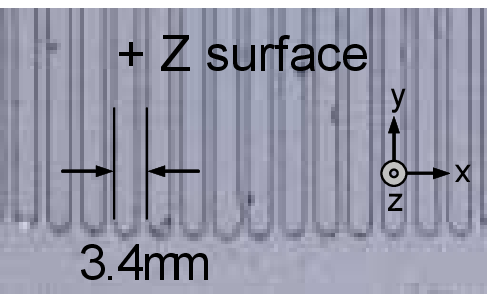}
}
\subfigure[]{
\includegraphics[width=8cm,clip]{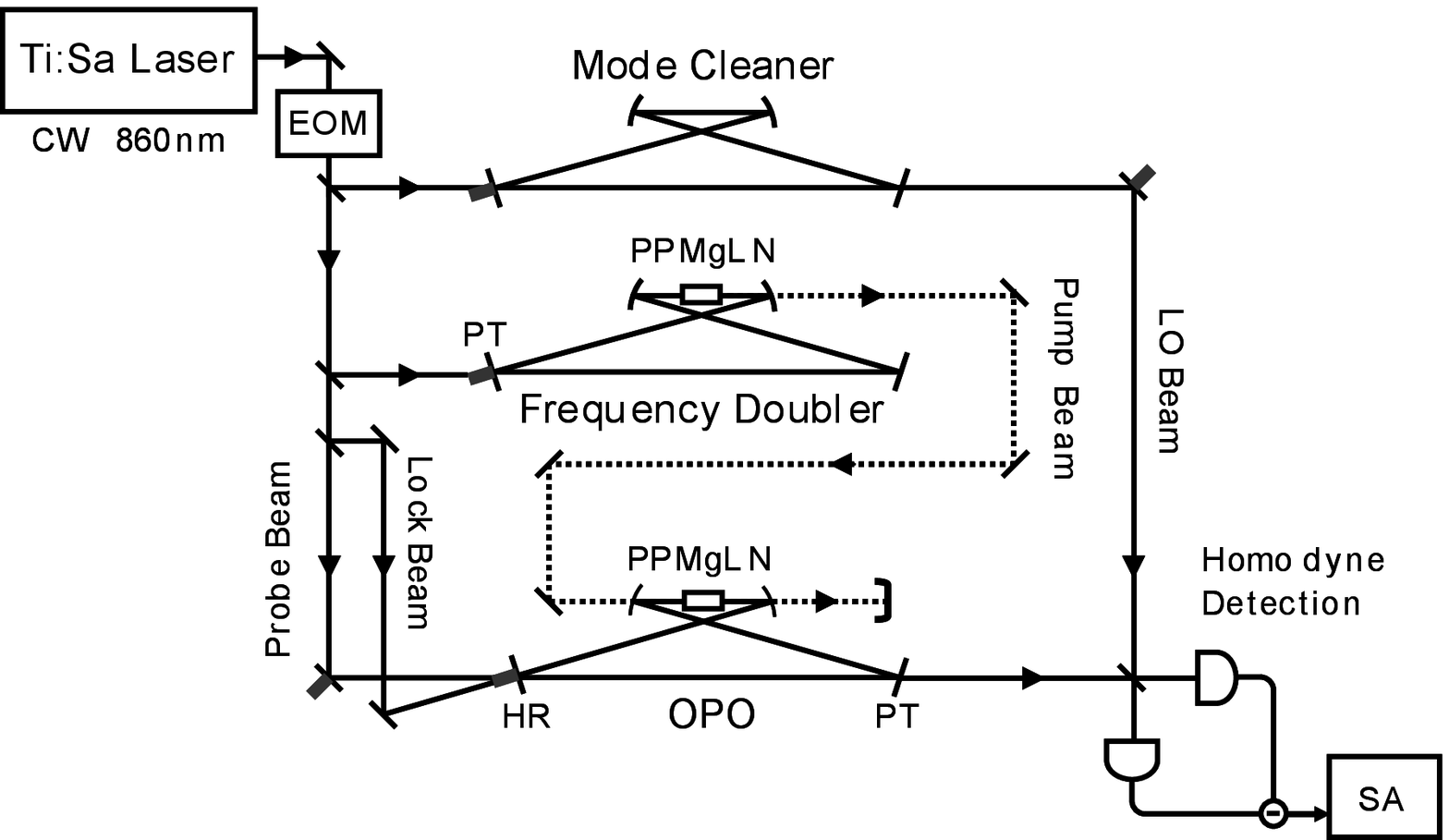}
}
\caption{(a)Observations of a periodically poled structure at +Z surface where an electrode was attached. (b)Schematic diagram of experimental setup. }
\label{FigSetup}
\end{figure}

In this work we focus on utilization of the largest optical coefficient $d_{\mathrm{33}}$ of MgLN crystal to improve the effective nonlinearity ${E}_{NL}$ which is essentially important to generate not only highly squeezed light but also second harmonic wave as a pump beam of the OPO. 
For that purpose we fabricated periodically poled MgO 5mol\% doped LiNbO$_{\mathrm{3}}$ (PPMgLN) crystals for utilizing the $d_{\mathrm{33}}$.
Periodically poled structure with 3.4 $\mu$m period was formed in MgLN crystal by temperature elevated field poling technique~\cite{Ishizuki03} as shown in Fig.~\ref{FigSetup}(a).
In consequence the effective nonlinearity of 0.043 ($\mathrm{W}^{-1}$) was realized.
We achieved the squeezing level of $-7.60 \pm0.15$dB at 860 nm by utilizing a subthreshold OPO with the PPMgLN crystal. 
We also generated second harmonic waves at 430 nm with conversion efficiency of 70\% by using the PPMgLN crystal inside an external cavity.
 
A schematic diagram of experimental setup is shown in Fig.~\ref{FigSetup}(b). 
We use a continuous-wave Ti:Sapphire laser at 860 nm as a light source.  
An optical system mainly consists of three cavities with a bow-tie configuration, an OPO, an optical frequency doubler to generate a pump beam of the OPO, and a mode cleaner for spatially filtering a local oscillator (LO) beam to make the same spatial mode with the OPO output. 
The 860 nm beam is phase modulated at 9.1 MHz by an electro-optic modulator (EOM) in order to lock the resonance of each cavity by conventional FM sideband locking technique~\cite{Drever83}. 
Both cavities of the OPO and the frequency doubler are designed as resonant at 860 nm and have two spherical mirrors whose radius of curvature is 50 mm and two flat mirrors.  
One of the flat mirrors has a partial transmittance (PT) at 860 nm and is used as a coupling mirror. 
A 9.5 mm and 8 mm-long PPMgLN crystals whose temperature is controlled around $50^\circ\mathrm{C}$ are placed between the two spherical mirrors of the OPO and the frequency doubler respectively.
The round trip length of the cavity is about 500 mm which yields the beam waist size of 21 $\mu$m in radius at the crystal center. 

\begin{figure}
\centering
\includegraphics[width=7.5cm,clip]{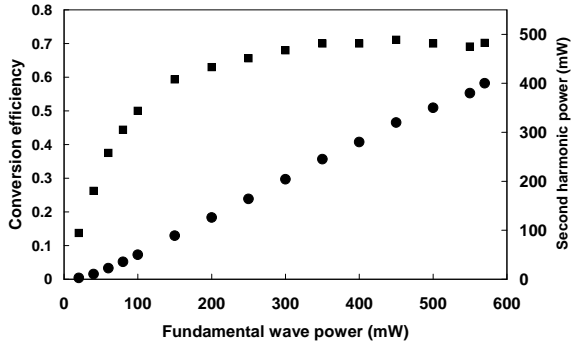}
\caption{A property of frequency doubler with the PPMgLN crystal. Circles and squares indicate the power of second harmonic wave at 430 nm and the conversion efficiency respectively.}
\label{FigDoubler}
\end{figure}

Firstly we characterized the effective nonlinearity ${E}_{NL}$ of the PPMgLN crystal with the OPO.
The ${E}_{NL}$ is defined as ${P_{\mathrm{2\omega}}}/P_{\mathrm{\omega}}^2$ ($\mathrm{W}^{-1}$) where $P_{\mathrm{\omega}}$ and $P_{\mathrm{2\omega}}$ are the power of fundamental wave and second harmonic respectively with a single pass frequency doubling. 
The result shows the ${E}_{NL}$ of 0.043 ($\mathrm{W}^{-1}$) which is two times larger than the previously reported ones for KNbO$_{\mathrm{3}}$ and PPKTP as shown in Table~\ref{table} under the similar focusing condition~\cite{Polzik92,Suzuki06}.
Theoretical estimation of the effective nonlinear optical coefficient, $d_{\textit{eff}}=({2}/{\pi})d_{\mathrm{33}}$, from the observed ${E}_{NL}$ by using the well-known theory of Boyd and Kleinman~\cite{Boyd68} yields 15 (pm/V) which  reasonably agrees with the previously reported value~\cite{Shoji97}.
A slight difference might be caused by imperfection of the periodically poled structure.
By utilizing high nonlinearity of the PPMgLN crystal we also constructed the frequency doubler with a coupling mirror whose transmittance is 0.10.
We achieved the second harmonic power of 400 mW from fundamental wave power of 570 mW which yielded the conversion efficiency of 70\% as shown in Fig.~\ref{FigDoubler}. 

Next we evaluated the intracavity loss $L$ of the OPO by injecting a weak coherent beam from the output coupler. 
The analysis shows the loss without a pump beam is 0.011, which is rather high compared with the previously reported crystals and might be caused by imperfection of the periodically poled structure, and increases up to 0.022 at the pump power of 350 mW probably due to BLIIRA and/or photo-refractive effect. 
The experimental results can be expressed as a following equation; $L=L_0+aP_{\mathrm{{2\omega}}}$, where $L_0$ is a passive loss and $a$ is a coefficient of pump induced losses and are calculated as $L_0=0.01236$ and $a=0.0246$ ($\mathrm{W}^{-1}$) respectively.

\begin{figure}
\centering
\subfigure[]{
\includegraphics[width=4cm,clip]{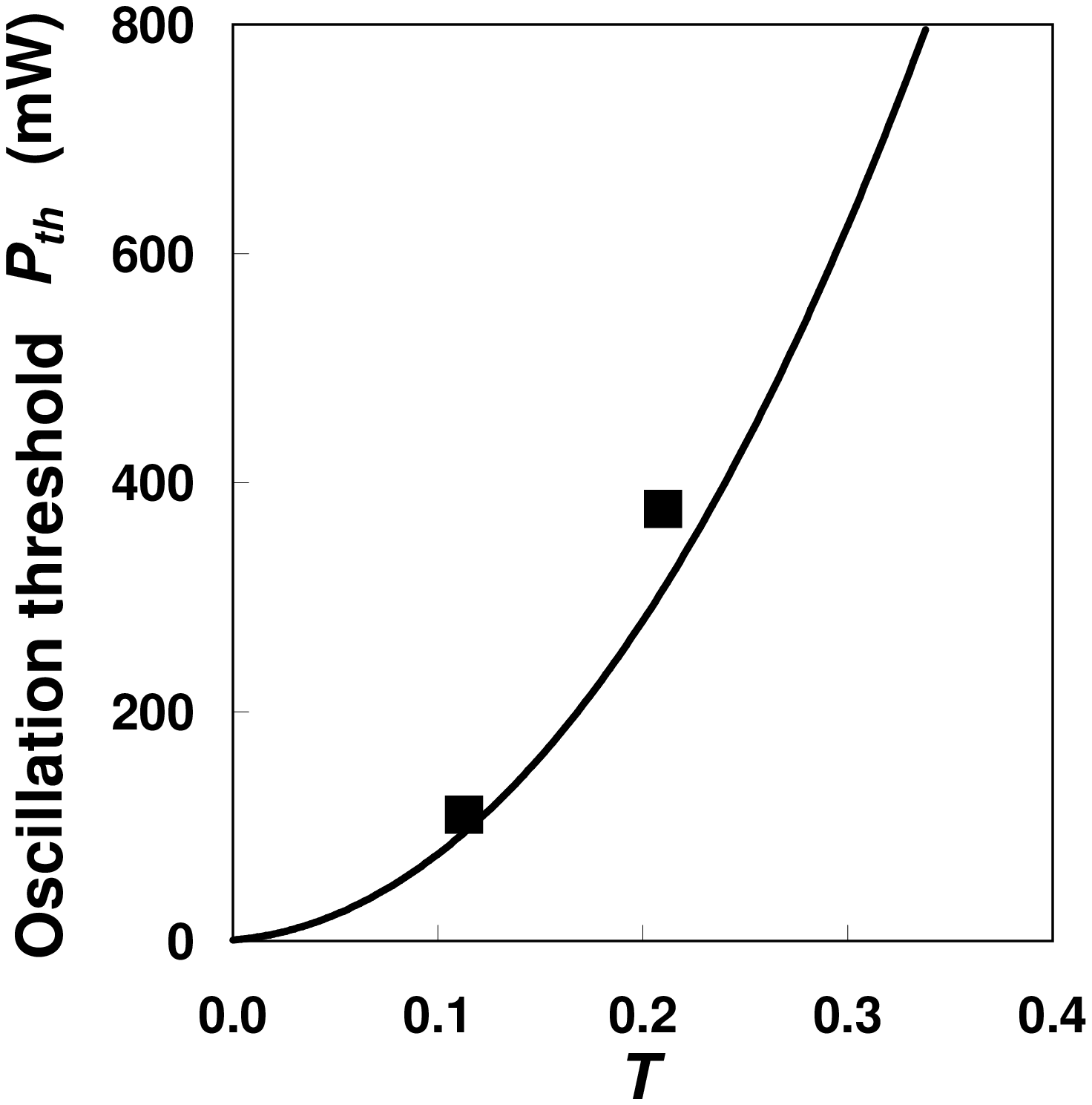}
}
\subfigure[]{
\includegraphics[width=4cm,clip]{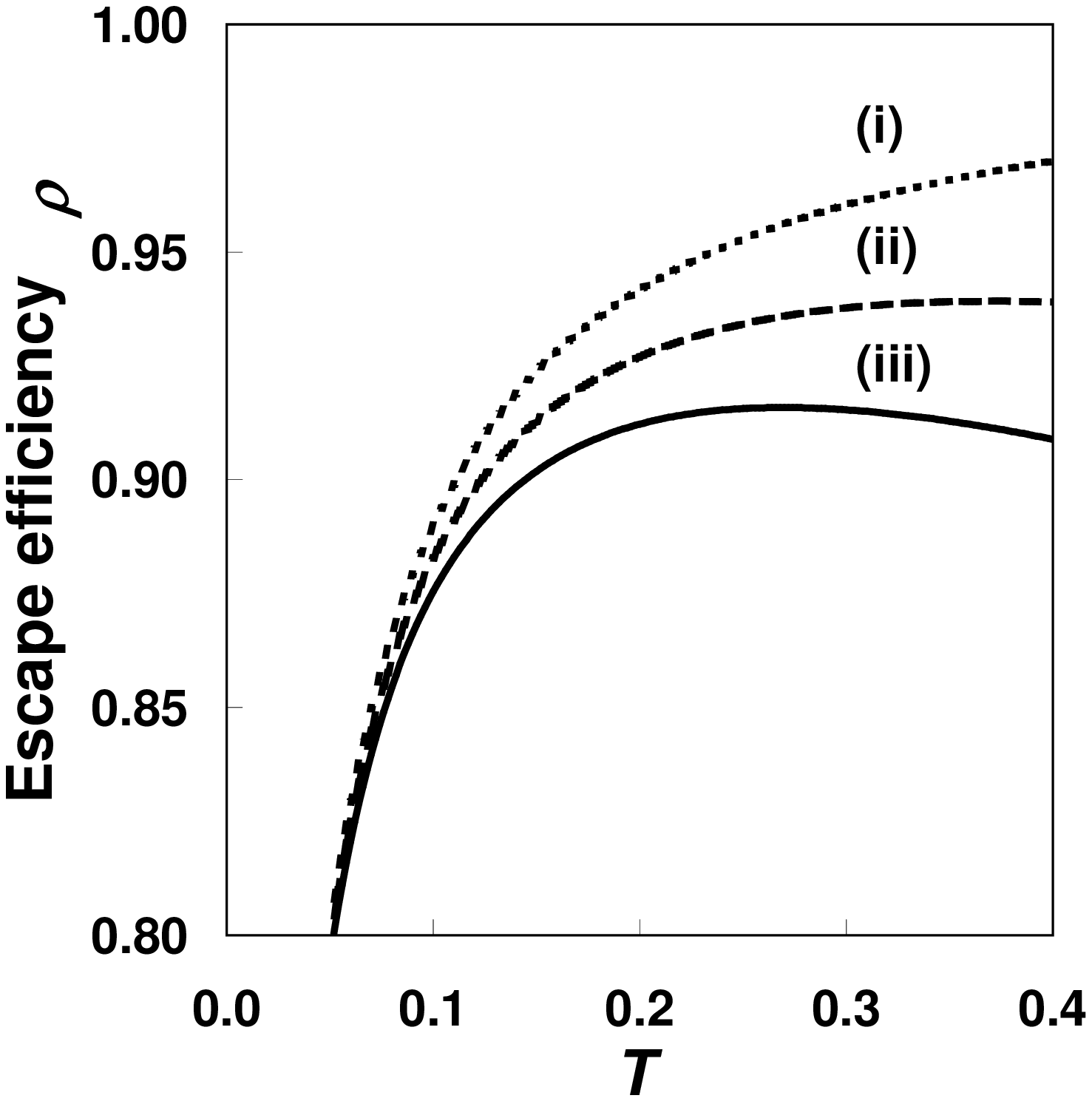}
}
\caption{(a)$T$ dependence of the Oscillation threshold. Solid line is calculation result and squares indicate experimental results. (b)Estimation of the escape efficiency at (i)$x$=0 (${P_{\mathrm{2\omega}}}=0$), (i\hspace{-.1em}i)$x$=0.7 (${P_{\mathrm{2\omega}}}\approx 0.5P_{th}$), and (i\hspace{-.1em}i\hspace{-.1em}i)$x$=1 (${P_{\mathrm{2\omega}}}=P_{th}$).}
\label{FigPthEscape}
\end{figure}

\begin{figure}
\centering
\subfigure[]{
\includegraphics[width=7.5cm,clip]{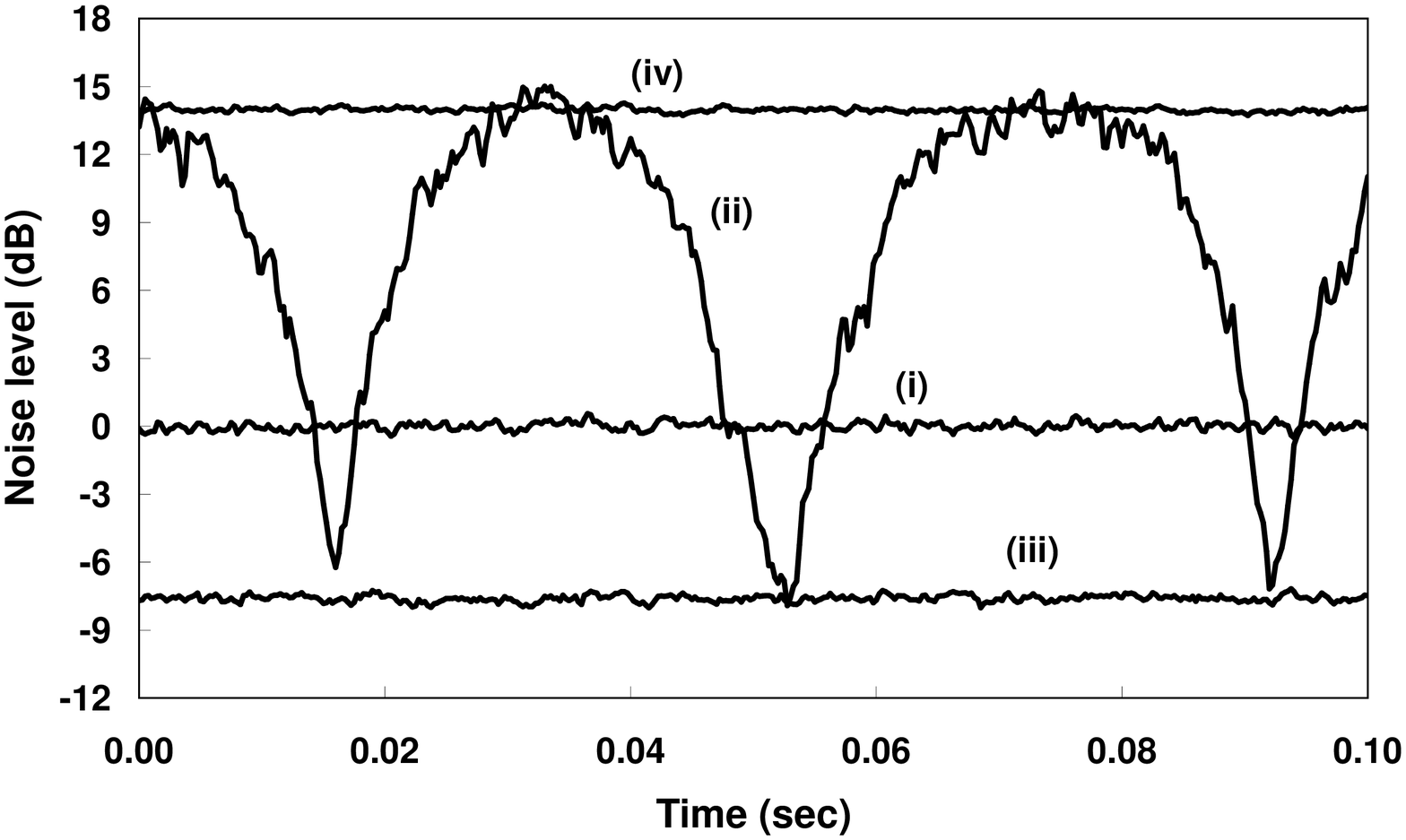}
}
\centering
\subfigure[]{
\includegraphics[width=7.5cm,clip]{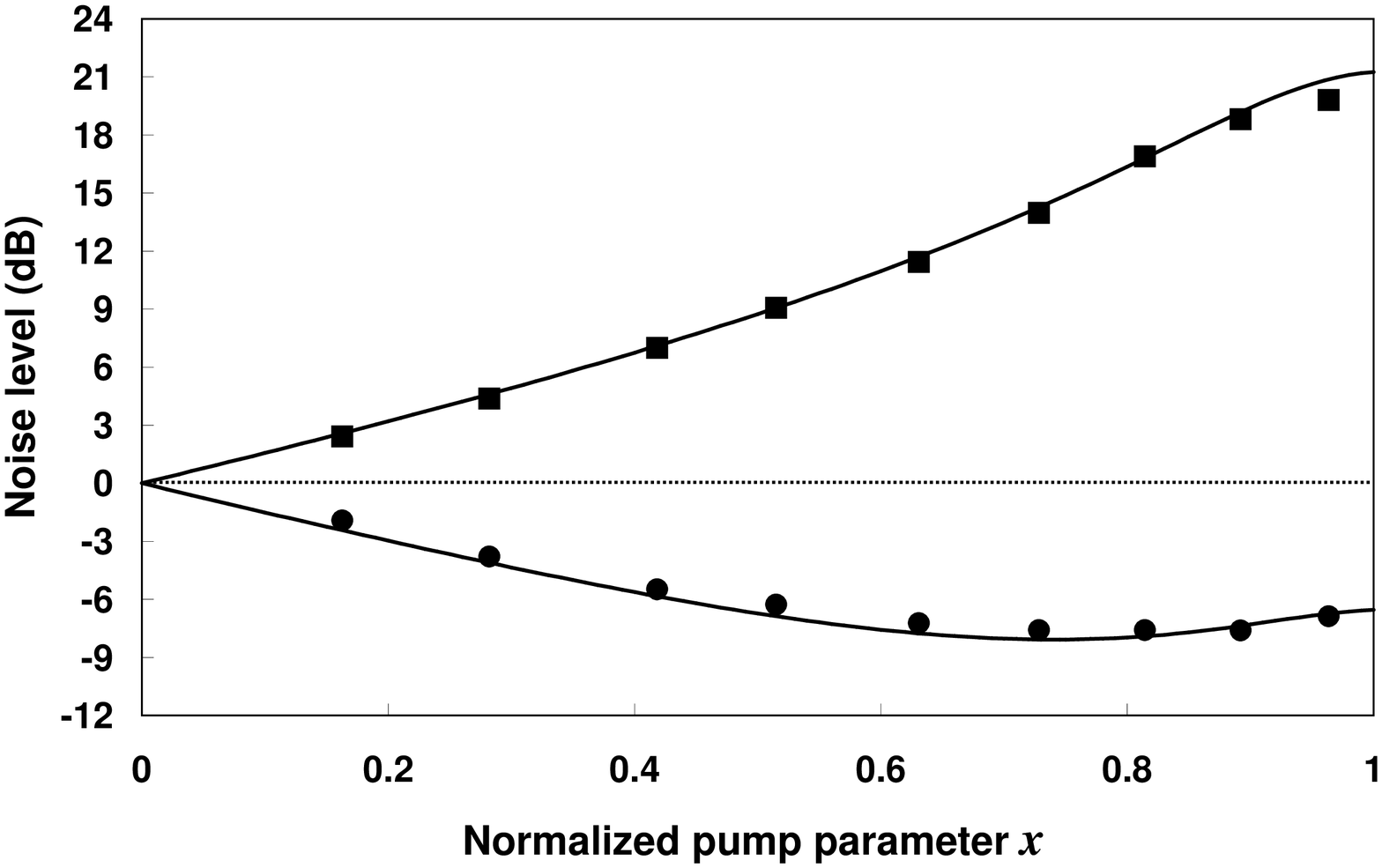}
}
\caption{(a)Noise level at pump power of 200mW. (i)Shot noise level, (i\hspace{-.1em}i)LO phase is scanned, (i\hspace{-.1em}i\hspace{-.1em}i)LO phase is locked at the squeezed quadrature, and (i\hspace{-.1em}v)LO phase is locked the antisqueezed quadrature. 
Traces (i), (i\hspace{-.1em}i\hspace{-.1em}i), and (i\hspace{-.1em}v) are averaged for 20 times. 
(b)Pump power dependence of the squeezing and antisqueezing levels. Circles/squares indicate the observed squeezing/antisqueezing levels. 
Solid curves indicate the results of theoretical calculation.}
\label{FigSQ}
\end{figure}

We measured the oscillation threshold $P_{th}$ of the OPO from the parametric gain when a weak coherent light (probe beam) is introduced from the highly reflective (HR) mirror of the OPO. 
The resonance of the OPO is locked by using a counter-propagating lock beam.  
When we used the output coupler with transmittance of 0.113, the $P_{th}$ was estimated as 110 mW which was much lower than previous works regardless of the large intracavity losses.
This is again because of the high nonlinearity of the PPMgLN crystal. 
High nonlinearity allows us to use high transmittance of output coupler to improve the escape efficiency. 
By using the relation of $L=L_0+aP_{\mathrm{{2\omega}}}$, we can calculate the values of $P_{th}=(T+L)^2/4E_{NL}$ and $\rho=T/(T+L)$ as a function of $T$, which are shown in Fig.~\ref{FigPthEscape}(a) and (b) respectively.
We estimated the $\rho$ at certain normalized pump parameter $x=\sqrt{P_{\mathrm{2\omega}}/P_{th}}$ in Fig.~\ref{FigPthEscape}(b). 
At a lower $x$ (Fig.~\ref{FigPthEscape}(b)(i)-(i\hspace{-.1em}i)), the $\rho$ improves monotonically by using a higher $T$.
However at a higher $x$ (Fig.~\ref{FigPthEscape}(b)(i\hspace{-.1em}i\hspace{-.1em}i)), the $\rho$ starts to degrade due to the increase of intracavity losses caused by high pump power. 
As a result there is an optimum $T$ which maximizes the escape efficiency at a higher $x$. 
So we decided to use the output coupler with 0.21 of $T$.
Although the $P_{th}$ is increased up to 377 mW (Fig.~\ref{FigPthEscape}(a)), nearly optimum value of the $\rho$ of 0.91 is expected at Fig.~\ref{FigPthEscape}(b)(i\hspace{-.1em}i\hspace{-.1em}i).
Another advantage of using a higher $T$ is broadening of a cavity bandwidth of the OPO which enables us to obtain the squeezing level at 2 MHz without degradation caused by detuning and it leads to avoid laser noise at lower frequency.

Fig.~\ref{FigSQ}(a) shows a typical result of the squeezing experiment at the pump power of 200 mW. 
The noise level is measured with a spectrum analyzer in zero span mode with the resolution band width of 30 kHz and the video bandwidth of 300 Hz. 
The shot noise level is 26 dB above a circuit noise of the homodyne detector. 
The observed squeezing level is $-7.60 \pm0.15$dB and antisqueezing level is $+13.97 \pm0.10$dB, respectively. 
We continued above experiment at several pump power and summarized in Fig.~\ref{FigSQ}(b).
The squeezing level saturates at the higher pump power level. 
To explain this result we calculated theoretical values of the squeezing and antisqueezing levels by the same analysis described in references~\cite{Suzuki06,Takeno07} with the intracavity loss of $L=L_0+aP_{\mathrm{{2\omega}}}$, a phase fluctuation of 1.5$^\circ$, and a homodyne detection efficiency of 0.968.
In Fig.~\ref{FigSQ}(b) theoretical curves agree well with the observed results. 
So the degradation of the squeezing level could be due to the pump induced losses and phase fluctuation of the LO beam.
If we could improve the intracavity losses less than 0.001 and suppress the phase fluctuation to 0.3$^\circ$ in future work, the squeezing level of -13 dB would be expected.

In conclusion, high nonlinearity was realized by fabricating a periodically poled MgO:LiNbO$_{\mathrm{3}}$ crystal for utilizing the large nonlinear optical coefficient $d_{\mathrm{33}}$.
We achieved the squeezing level of $-7.60 \pm0.15$dB and antisqueezing level of $+13.97 \pm0.10$dB respectively with the PPMgLN crystal. 
We also generated 400 mW of second harmonic waves at 430 nm with 70\% of conversion efficiency by using the PPMgLN crystal.

This work is partly supported by SCF, GIA, G-COE, and PFN commissioned by the MEXT, RFOST, and SCOPE of the MIC.

\thispagestyle{empty}

\renewcommand{\refname}{\vspace{-1.0cm}}

\end{document}